# Near infrared emission bands of solid Nitrogen

E.V. Savchenko, I.V. Khyzhniy, S.A. Uyutnov, M.A. Bludov,

B. Verkin Institute for Low Temperature Physics & Engineering NASU, Kharkiv 61103, Ukraine

**Abstract**

New results on the study of radiation effects in solid nitrogen and $N_2$-doped Ne matrix are presented with a focus on the so-called γ-line origin. The irradiation was carried out in dc regime with an electron beam of subthreshold energy. Relaxation dynamics was monitored by emission spectroscopy – cathodoluminescence (CL) and nonstationary luminescence (NsL), along with current activation spectroscopy. Thermally stimulated luminescence (TSL) and exoelectron emission (TSEE) of pure nitrogen and $N_2$ in the Ne matrix were measured in a correlated manner. Three emission bands were recorded in the NIR CL spectra of solid $N_2$: 794, 802, and 810 nm. The band at 810 nm was detected for the first time. These three bands are characterized by similar behavior and form molecular series with spacing between adjacent vibrational energy levels of the ground state of 125 cm$^{-1}$ and 123 cm$^{-1}$. These data cast doubt on the recently made assumption that the γ–line is attributed to the emission of the nitrogen anion N$^-$ [19]. The processes of an electron attachment and neutralization of positively charged species are discussed. It has been established that the γ-line in the TSL spectra of pure nitrogen and $N_2$-doped Ne matrix correlates with TSEE currents and recombination emission of O$^+$, $N_2^+$, and $N_4^+$ ions, which indicates its connection with the neutralization reaction. The measurement of NsL supported this conclusion. A new possible assignment of the γ-line and its satellites to the emission of tetranitrogen $N_4$ is discussed.

**Key words**: solid nitrogen, nitrogen anions and cations, emission spectroscopy, activation spectroscopy, neutralization reactions, tetranitrogen.

**Introduction**

Nitrogen solids gained general recognition as classical model molecular crystals which attract much attention in diverse fields of science – physics and chemistry of interstellar and solar systems [1-8], material science, specifically, the problem of polynitrogen compounds considered as environment-friendly high energy-density materials (HEDM) [9-12]. Energy storage, transformation, and its release are the focus of studies and among the methods used spectroscopy is one of the most effective. Optical spectroscopy of solid nitrogen has a long story and its recent trends were reviewed in [13]. Despite an impressive progress there are still questions under discussion and the most long-standing one concerns identification of, the so-called γ-line, situated in the near infrared (NIR) range. This line was observed both during the condensation of samples from the discharge and their subsequent heating [14-19], and during the irradiation of solid nitrogen films with electrons [20-23] and fast atoms [24]. The most detailed study of the γ-line was performed using the technique of growing impurity helium condensates (IHCs) by injecting a gas jet of helium with impurities ($N_2$ and mixtures of $N_2$ with Ne or Ar or Kr), which had preliminary passed through the discharge, into superfluid helium (HeII) [19]. These IHCs represented porous gel-like structures formed by impurity clusters with the characteristic size of 3–10 nm. The γ-line position depended on the type of impurity forming the nanoclusters and for $N_2$, Ar and Ne varied around 793 nm according [19]. Its position in Kr clusters was about 795.5 nm. In the most intense spectrum of nitrogen nanoclusters the authors observed a red-shifted satellite at 801.8 nm. These data are consistent with the results obtained by another method for production of excited particles, i.e., irradiation of condensed films with electrons [20]. Comparative study of thermally stimulated luminescence (TSL) of nanoclusters and solid nitrogen films pre-irradiated with an electron beam [23] demonstrated a similarity of spectra and revealed thermally stimulated emission of electrons (TSEE) from both kind of samples. It was found that the yield of TSEE from the nitrogen film correlates with the TSL recorded on the γ-line and on the α-group connected with the doubly forbidden $^2D \rightarrow {}^4S$ transition of N atom. Correlation of the TSEE with the total yield of TSL was established for nitrogen nanoclusters. In subsequent work [19], the correlation of three emissions, TSEE, TSL(γ), and TSL(α) during thermal destruction of a nitrogen-helium sample was confirmed. The similarity of the glow curves detected on the γ-line and on the α-group which correlated with the TSEE yield for both kind of samples implies a similar mechanism for the population of emitting states and points to an important role of electrons in this process. Radiation effects of fast-particle bombardment of solid nitrogen were studied employing technique of fast-atom-bombardment mass spectrometry (FAB MS), ESR and optical spectroscopy [24]. The mass-spectrum of the ejected material contained $N^+$, $N_2^+$, $N_3^+(N_2)_n$, and $N_4^+(N_2)_n$, as well as a trace of $N_3^-$. ESR and optical spectra of $N_2$ matrices bombarded either with Ne or with Ar ions and/or neutrals showed the presence of nitrogen atoms $N(^4S)$, $N(^2D)$, the radical $N_3(X\ ^2\Pi_g)$, and the anion $N_3^-(X\ ^1\Sigma_g^+)$. The emission spectrum recorded during bombardment with a Ne + Ne$^+$ beam of keV-energy contained the γ-line at 793 nm with a short lifetime (<1 s). The authors following [14] attributed this emission to nitrogen atoms. In [22], despite the fact that this line did not shift upon $O_2^{18}$ substitution the authors assigned it to the (0, 0) transition of the atmospheric bands of oxygen ($A\ ^1\Sigma_g^+ \rightarrow X\ ^3\Sigma_g^-$) which lies at 759.4 nm in the gas phase. In a recent article [19], a new hypothesis was put forward for the origin of the γ-line as the emission of the nitrogen anion $N^-$. It was suggested that the γ-line appears as a result of the interaction of mobile electrons and metastable $N(^2D)$ atoms in solid matrices resulting in an electron attachment with formation of the excited state of the nitrogen anion $N^-$:

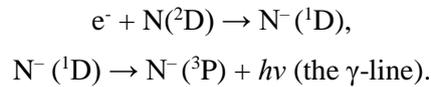

$$e^- + N(^2D) \rightarrow N^-(^1D),$$

$$N^-(^1D) \rightarrow N^-(^3P) + h\nu \text{ (the γ-line)}.$$

However, this hypothesis did not explain an origin of the red satellite of the γ-line observed in a number of studies, e.g. [17, 19, 20].

To test this hypothesis we performed new experiments on radiation effects in solid nitrogen and $N_2$-doped Ne matrix employing complex of emission spectroscopy methods: cathodoluminescence (CL), nonstationary luminescence (NsL), along with optical and current activation spectroscopy. In the NIR CL spectra of solid $N_2$ three emission bands were recorded: at 794, 802 and 810 nm. The band at 810 nm was detected for the first time. These three bands are characterized by similar behavior that call into question the identification of the γ-line as the emission of nitrogen anion $N^-$ [19]. The processes of an electron attachment and neutralization of positively charged species are discussed. It was shown that the γ-line in the TSL spectra of pure nitrogen and $N_2$-doped Ne matrix correlates with the TSEE currents indicating its connection with neutralization reaction. The measurements of NsL supported this conclusion. A new possible assignment of the γ-line and its satellites to the emission of tetranitrogen $N_4$ is discussed.

**Experimental**

Basic principles of the experimental approach to study radiation effects in frozen gases and specifically in solid nitrogen have been described in detail in sect. 7 of ref. [26] and in [13]. Here, we give a brief account of the procedures. The experiments were performed with pure solid nitrogen and with nitrogen-doped Ne matrices in a high-vacuum chamber with a base pressure of P<10$^{-7}$torr. The films of 25 μm thick were grown by deposition of pure room temperature N$_2$ gas or mixture N$_2$/Ne of chosen concentration onto a cold oxygen-free copper substrate mounted on a liquid He cryostat finger. The grown films were transparent with a smooth surface. The open surface provides the possibility of spectral measurements in a wide range (from VUV up to NIR) and detection of stimulated electron emission. The irradiation was performed in dc regime with an electron beam of subthreshold energy $E$=500 eV to exclude the knock-on defect formation and sputtering. The beam covered the icy film with an area of 1 cm$^2$. The sample heating under electron beam did not exceed 0.4 K. The luminescence spectra were recorded simultaneously in VUV, visible and NIR range (from 50 nm up to 1000 nm) using two ports of the experimental chamber. Note, that spectra were not corrected for the spectral sensitivity of the optical system. An additional chamber port was used for visual control and allowed experiments with photoexcitation.

On completing the exposure we detected an "afteremission" current. This phenomenon was observed in [27] which reported first detection of TSEE from solid nitrogen pre-irradiated with an electron beam. Measurements of thermally stimulated relaxation emissions TSL and TSEE from solid nitrogen were started when the "afteremission" current and afterglow had decayed to essentially zero. In these experiments we used heating with a constant rate of 5 K min$^{-1}$. The required temperature measured with a Si sensor was set using the heater. The preset heating rate of the sample was controlled by the program. Released from shallow traps electrons being promoted to the conduction band either neutralize positively charged centers or escape from the film yielding TSEE current. Stimulated currents were detected with an electrode kept at a small positive potential $V_F$=+9 V and connected to the current amplifier. Due to high mobility of electrons in solid nitrogen in the range of α-phase existence [27] and their relatively high mobility in solid Ne [28] TSEE measurements provide information not only on surface-related processes but also on the processes occurring in deeper layers of the films and on charge accumulation. In view of the sensitivity of TSEE, TSL, and other phenomena to the sample structure, impurity concentration, and other variables, these measurements were performed in a correlated manner on the same sample. The combination of optical and current activation spectroscopy methods made it possible to track various relaxation channels and obtain information about both charged and neutral particles. In addition, we used the method of nonstationary luminescence (NsL) developed by our group [29]. This two-stage method was introduced to probe charged species and neutralization reactions. The main idea of this method is to create conditions when localized charge carriers of different signs come into play. For the purpose at the first stage the charged species of interest are generated with an electron beam of a high current density. At the second stage the species under study are probed with a low-density beam on heating. In doing so, we reveal the contribution of the neutralization reaction driven by the released electrons to the NsL measured on the γ –line. The experiments with pure nitrogen were performed in the range of the a-phase existence. In the Ne matrix the temperature range studied was 5-15 K.

**Results and discussion**

Fig. 1 illustrates the typical CL spectrum of solid nitrogen in the VUV, visible and NIR range.

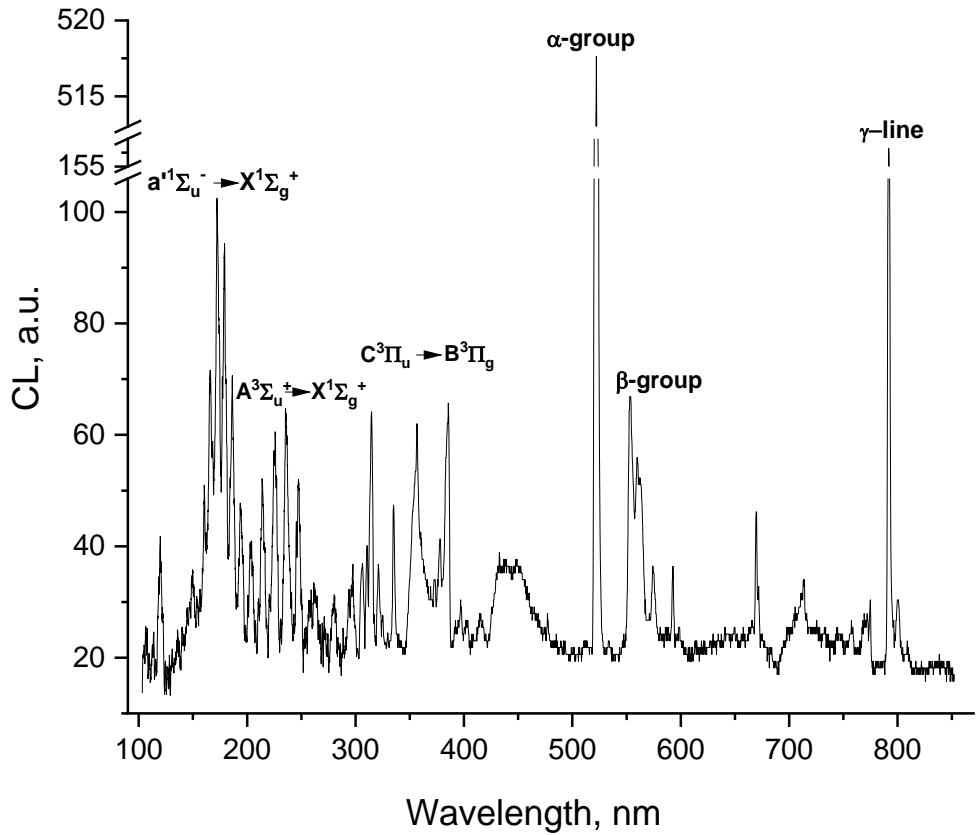

Fig. 1 Cathodoluminescence spectrum of solid nitrogen recorded upon excitation with a 500 eV electron beam.

The most intense bands in the VUV range of the spectrum of pure nitrogen are the singlet molecular progression $a'^1\Sigma_u^- \rightarrow X^1\Sigma_g^+$ and the Vegard-Kaplan intercombination transition $A^3\Sigma_u^+ \rightarrow X^1\Sigma_g^+$. The positions of the vibrational bands of both progressions are in good agreement with those measured earlier and reported in [13]. All the vibrational bands of both progressions show a matrix shift towards lower energy in comparison with the gas phase spectra indicating the bulk origin of the emitting species. The shift for the $a'^1\Sigma_u^- \rightarrow X^1\Sigma_g^+$ progression is 450 cm$^{-1}$ and 330 cm$^{-1}$ for the $A^3\Sigma_u^+ \rightarrow X^1\Sigma_g^+$ one. Spectral line at 120 nm stems from the $3s^4P_{5/2-1-2} \rightarrow 2p^3\ ^4S_{3/2}$ transitions. The absence of matrix shift indicates that this line is emitted by excited N* atoms which have left the sample surface. A possible mechanism of the excited N* atom formation via dissociative recombination of $N_3^+$ is considered in [30]. The spectrum in visible and NIR range is dominated by the α-group corresponding to the emission of N atoms initiated by matrix phonons. The α'-group represents the simultaneous vibrational excitation of $N_2$ molecule ($v=0 \rightarrow v=1$) with the atomic $^2D \rightarrow ^4S$ transition [31]. The β-group stems from an oxygen impurity and corresponds to the $O(^1S) \rightarrow O(^3P)$ transition of O atom. In the near UV range the emission of the second positive system – the transitions between $C^3\Pi_u$ and $B^3\Pi_g$ excited molecular states, was registered against the background of a wide band at 360 nm, which will be discussed later. The second positive system predominates in thin samples and at low electron beam energy. The distinctive feature of this emission is the coincidence of the observed bands with those detected in the gas phase spectra. As it was shown in [32] this emission belongs to $N_2$ molecules desorbing in the excited $C^3\Pi_u$ state. The peak near 388 nm stems from the impurity center CN ($B^2\Sigma^+ \rightarrow X^2\Sigma^+$ 0-0 line). Some lines above 600 nm belong to the second spectral order. In the NIR range the γ-line with satellites was detected. The enlarged part of the spectrum in the range of the γ-line is shown in Fig. 2.

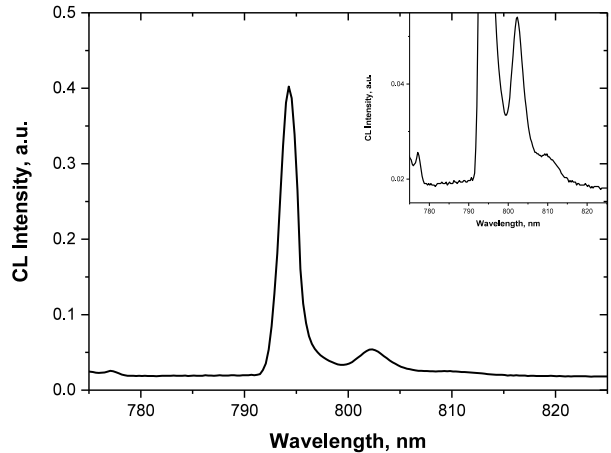

Fig.2 Cathodoluminescence spectrum of solid nitrogen in the range of γ-line

The bands at 794 and 802 nm are close to those observed in films of solid nitrogen bombarded by 20 keV electrons [20] and upon destruction of the impurity helium condensates grown from discharge [19]. The band at 810 nm was detected for the first time. These three bands are characterized by similar behavior and form molecular series with spacing between adjacent vibrational energy levels of the ground state of 125 cm$^{-1}$ and 123 cm$^{-1}$. These findings call into question the identification of the γ–line as the emission of nitrogen anion N$^-$.

Let us consider in more detail the data on activation spectroscopy. Since TSL can occur both in the recombination of charged particles and in the recombination of neutrals, it must be used in combination with the methods of current activation spectroscopy, e.g. TSEE. The electrons released from the traps can contribute to two relaxation channels: (i) electron attachment to particles with positive electron affinity and (ii) electron recombination with positively charged species. In the case (i), anions are formed in the ground state if an electron is attached to an atom or molecule in the ground state. If it is the case, no contribution to the TSL spectra is expected. Anions in the ground state can be traced using absorption spectroscopy or other methods. So, formation of N$_3^-$ was detected in the early study [24] under fast atom bombardment using absorption spectroscopy, ESR and mass spectrometry. N$_3^-$ anion was identified in ices warmed to 35 K [33]. One more technique was used to detect N$_3^-$ – photon stimulated exoelectron emission (PSEE) [34, 35]. In these experiments a sharp surge of PSEE current was recorded upon switching 2.76 eV photon flux that is the energy equal to the electron affinity χ of N$_3$ (χ=2.76 eV [36]). The only possibility to observe anions in TSL is the case when the positive values of electron affinity exceed the energy of the anion excited state formation. This case was considered in [19] for the $^2$D nitrogen atom. The electron affinity χ of the ground state N atom is very small χ=0.0725 eV [37]. However, for the $^2$D nitrogen atom the situation is completely different. The electron affinity χ for this state increases by more than an order of magnitude [38] that allows the formation of the nitrogen anion in the excited states N$^-$($^1$D) and N$^-$($^1$S) as suggested in [19]. Such a mechanism for the formation of an excited anion and the observation of the γ-line suggests the simultaneous presence of excited nitrogen atoms in the $^2$D state and electrons. The well-known afterglow of solid nitrogen is characterized by a long lifetime τ=37 s [39]. The observed afteremission current is also characterized by a long lifetime. Its curve can be fitted by a sum of two exponents with τ$_1$=37 s and τ$_2$=216 s. In other words, the above condition seems to be satisfied and one can expect the appearance of a γ line in the afterglow spectrum. The lifetime of the γ-line turned out to be equal to τ<1 s [19, 24], and only a feature associated with the $^2$D

→ $^4S$ transition of the nitrogen atom (α-group) was observed in the afterglow spectrum, at least within the sensitivity of our device. This observation agrees with the conclusion made in [14]. However, the absence of the γ-line in the afterglow spectrum can be explained if take into account that only slow thermalized electrons are capable to come into the electron attachment reaction. High and long lasting afteremission currents from negatively charged nitrogen films [35] assume only minor contribution of slow electrons in the process. The same statement is true for the neutralization processes. Note that, in clusters formed by electron attachment to helium nanodroplets doped with $N_2$ neither $N^-$ nor $N_2^-$ were found in [40]. The authors observed only odd $N_m^-$ ion series with $3 \leq m < 140$.

Dynamics of TSL spectra of a solid nitrogen film recorded in the visible and NIR ranges is shown in Fig. 3.

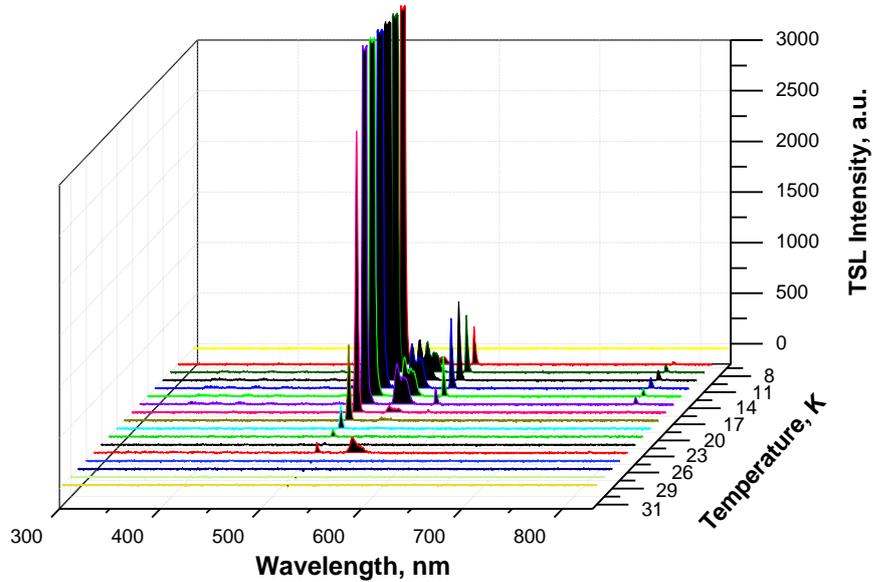

Fig.3 TSL of nitrogen film pre-irradiated with an electron beam.

It was found that the TSL detected at the γ-line correlates not only with the TSL recorded at the α-band, connected with the $^2D \rightarrow ^4S$ transition of N atom as it was reported in [23, 19]. Such a correlation was also found for the TSL recorded at the ($^1S \rightarrow ^3P$) transition of O atom presented in Fig. 4. It is noteworthy that even for the O atom, characterized by a very high positive electron affinity $\chi = 1.46$ eV [41], the neutralization channel of relaxation is predominant. One more correlating with the TSEE current TSL recorded at the $a'^1\Sigma_u^- \rightarrow X^1\Sigma_g^+$ transition of molecular nitrogen was attributed to the neutralization of the tetranitrogen cation $N_4^+$ [34, 42]. Correlation of all these partial yields of the TSL with TSEE current indicates connection of these emissions with the neutralization reactions stimulated by electron detrapping.

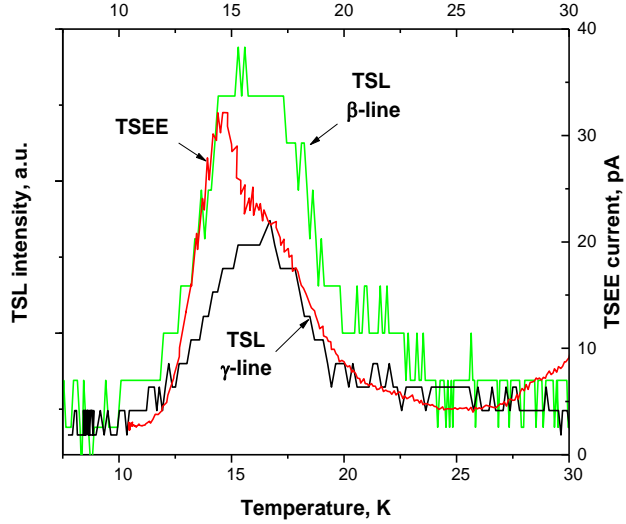

Fig. 4. Comparison of the TSL recorded at the γ-line and β-group with the TSEE yield.

No shift was observed between the yields of TSL and TSEE due to the negative electron affinity of solid nitrogen (-1.8 eV) [43], which resulted in the absence of a barrier to electron escape. Interestingly, different group of electron traps are involved in the neutralization of species responsible for TSL at the γ and β-bands – shallower ones for O centers and deeper ones for the γ-line emitters. This may be caused by the contamination of the surface layers of the sample because of the not very high vacuum in the chamber (P~$10^{-7}$ torr). An additional confirmation that neutralization reactions are the basis of the TSL process is a comparison of the TSEE yields with the NsL yields recorded at the same transition. The correlation between NsL recorded at the γ-line and the TSEE yield is separately illustrated by Fig. 5.

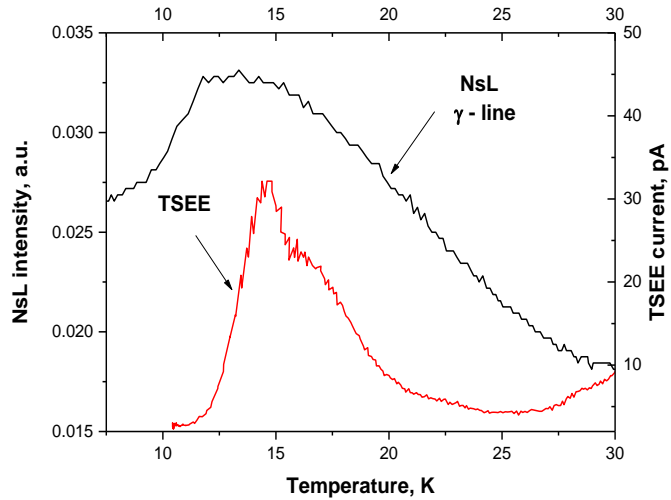

Fig.5 NsL at the γ-line measured upon heating under beam and TSEE yield of nitrogen film pre-irradiated with an electron beam.

The most likely recombination channels of positively charged nitrogen compounds with electrons are discussed below.

$$N^+ + e^- \rightarrow N(^2D) + \Delta E \rightarrow N + h\nu_1 \text{ (α-group)} + \Delta E_1 \quad (1)$$

Holes ($N_2^+$) in solid nitrogen are characterized by very low mobility [44] which suggests their self-trapping with the formation of the $N_4^+$ complex. Neutralization of the $N_3^+$ cation can proceed via two-body and three-body channels [45]:

$$N_3^+ + e^- \rightarrow N^* + N_2 + \Delta E \rightarrow N + N_2 + h\nu_1/h\nu_2 + \Delta E_2 \quad (2)$$

$$N_3^+ + e^- \rightarrow N + N + N + \Delta E_3 \quad (3)$$

The energy release in the two-body channel (2) comprises $\Delta E_2 = 10.5 eV$ if all products are formed in the ground state [45]. A large amount of energy released in channel (2) permits formation of metastable $N(^2D)$ atoms via this channel and creates preconditions for the dissociative recombination with desorption of $N^*$ atoms in the $3s\ ^4P_{5/2-1/2}$ state [30]. The second channel of dissociative recombination of $N_3^+$ with decay into three nitrogen atoms is energetically unfavorable.

The cation $N_4^+$ can recombine with electron by scenario „cage exit" with dissociation of the transient products $N_4^{**}$ into $N_2^*$ which relax radiatively into the ground state:

$$N_4^+ + e^- \rightarrow N_4^{**} \rightarrow N_2^* + N_2^* + \Delta E \rightarrow N_2 + N_2 + 2h\nu_3 + \Delta E_4 \quad (4)$$

However, there could be another relaxation scenario, so-called „cage effect" [46, 47]. The excited $N_2^*$ molecule after dissociation could relax with a ground state $N_2$ molecule from the surroundings to form $N_4^*$ and then $N_4$.

$$N_4^+ + e^- \rightarrow N_4^{**} \rightarrow N_4^* \rightarrow N_2^* + N_2 + \Delta E \rightarrow N_4^* \rightarrow N_4 + h\nu_4 + \Delta E_5 \quad (5)$$

The realization of „cage exit" or „cage effect" scenarios is determined by the presence of barriers to dissociation on the potential energy curves.

A likely way to obtain neutral $N_4$ is to start with the $N_4^+$ cation due to its relative stability and ease of preparation [9]. An appropriate experimental technique used for such route is the neutralization reionization mass spectrometry (NRMS). There have been several studies using this method that have reported the formation of neutral $N_4$, e.g. [48-50]. However, these experiments do not make it possible to determine either the electronic state of this compound or its structure. In fact, activation spectroscopy methods based on neutralization reactions can also be considered as an alternative route to generate $N_4$ neutral. Moreover, this technique involves carrying out experiments with solid samples at low temperatures $T \ll T_{room}$, which creates favorable conditions for preventing the dissociation of neutralization products. It is known that $N_4^+$ cation is formed in Ne matrix doped with nitrogen under electron excitation via the ion-neutral reaction $N_2 + N_2^+$ [51]. The ESR parameters obtained in this experiment are compatible with a linear centro-symmetric radical $N_4^+(D_{\infty h})$ with a $^2\Sigma_u$ ground electronic state. This assignment was confirmed by the study of IR absorption of a neon matrix doped with nitrogen [52]. With this in mind we performed measurements of the TSL spectra of nitrogen-doped Ne matrix pre-irradiated with a 500 eV electron beam. Dynamics of the TSL spectra is shown in Fig. 6. The TSL spectra are dominated by the progression $A^3\Sigma_u^+ \rightarrow X^1\Sigma_g^+$, which is much more intense than in the TSL of pure nitrogen and the α-group. There are also seen the β-group and the γ-line.

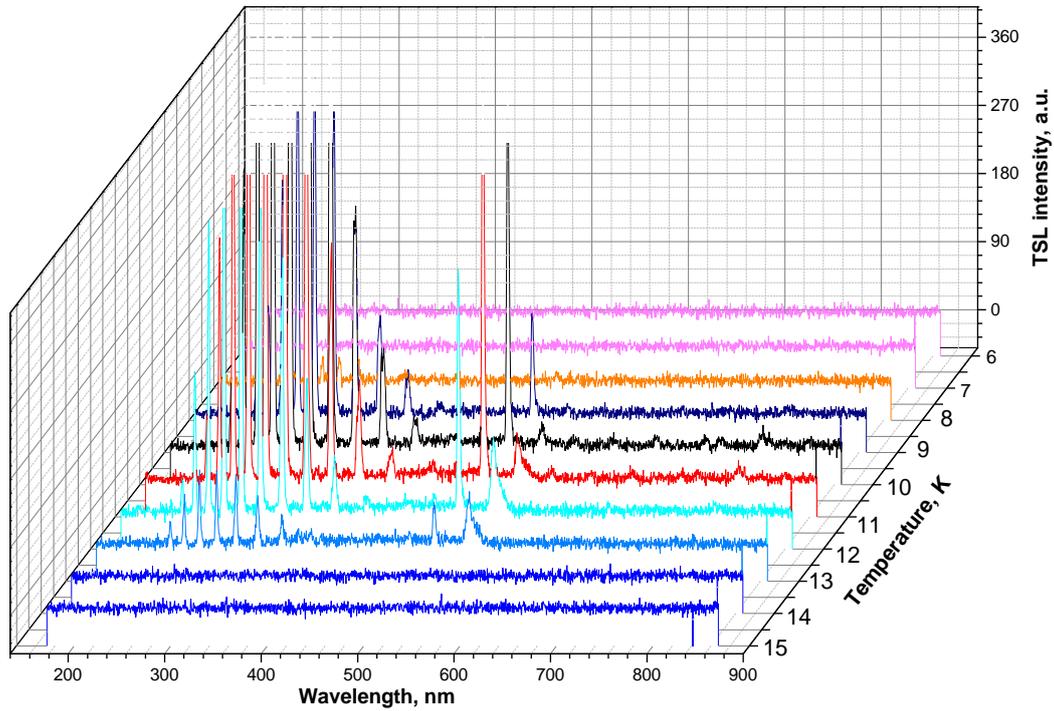

Fig. 6. The TSL spectra of Ne matrix doped with C=1% of $N_2$

The TSL at the γ-line measured simultaneously with the TSEE yield is presented separately in Fig. 7. As in the case of nitrogen, due to the negative electron affinity of Ne (-1.3 eV [28]), there is no shift of TSEE relative to TSL. As in pure nitrogen, TSL at the γ-line correlates with TSL measured at the α and β-bands. We also found the correlation with TSL at the Vegard-Kaplan molecular transition. The corresponding TSL curve for the 0-9 band of the $A^3\Sigma_u^+ \rightarrow X^1\Sigma_g^+$ progression is shown in Fig. 7. All these emissions correspond to the recombination luminescence of cations $N^+$, $O^+$, $N_2^+$ formed during irradiation. A clear correlation of these relaxation emissions points to the connection of the γ-line with the neutralization reaction, the most likely, of the $N_4^+$ tetranitrogen cation.

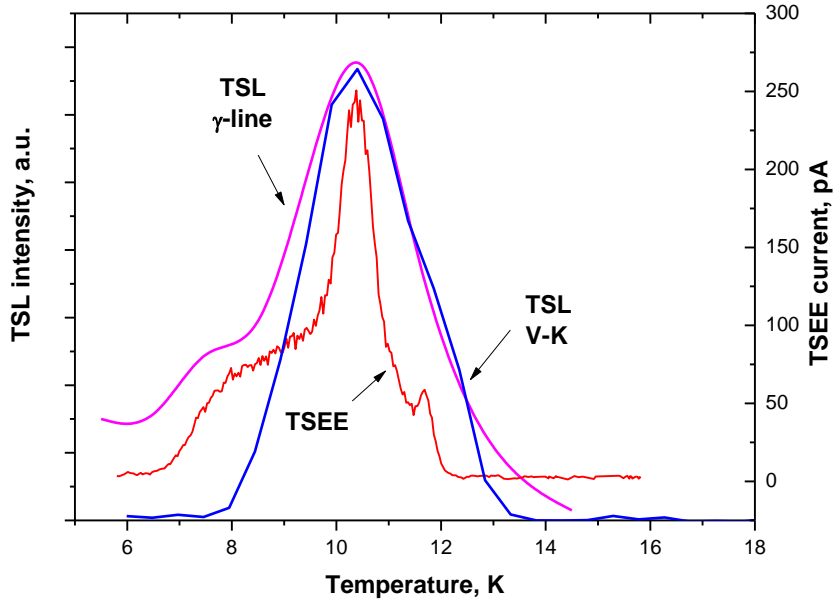

Fig. 7. The TSL of $N_2$-doped Ne matrix taken at the γ-line and on 0-9 band of the $A^3\Sigma_u^+ \rightarrow X^1\Sigma_g^+$ progression both recorded simultaneously with the TSEE current. TSL intensities were normalized to facilitate comparison.

Note, that due to the wide band gap $E_g$=21.58 eV [28], the Ne matrix is extremely convenient for the production and storage of various ionic species, many of which were extensively studied by the matrix isolation technique (e.g. Ref. [53] and references therein). Considering that the ionization potentials of the N atom, $N_2$ molecule, and $(N_2)_2$ dimer are 14.53, 15.58, and 14.6 eV, respectively, [9,54] one could expect the formation of the corresponding $N^+$, $N_2^+$, and $N_4^+$ cations. These species remain stable as long as the matrix is maintained cold and electrons are trapped in shallow (defects of structure) and deep (species with positive χ) traps. In addition to growth defects, electron-induced defects are created in the matrix during irradiation [55]. Heating of the pre-irradiated matrix triggers the release of electrons and relaxation processes. Ions $N^+$ recombine with electrons by reaction (1) contributing to the α-band. This emission can also emerge upon the dissociative recombination of the $N_3^+$ ion by reaction (2). This band is clearly seen in the TSL spectra in Fig. 6. The presence of an oxygen impurity leads to the dissociation of the $O_2$ molecule during irradiation and the appearance of an O atom, which is then ionized in the matrix. Its recombination proceeds according to reaction (6).

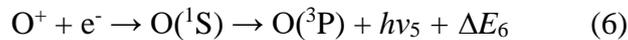
$$O^+ + e^- \rightarrow O(^1S) \rightarrow O(^3P) + h\nu_5 + \Delta E_6 \quad (6)$$

The TSL yield of the $O(^1S)$ emission for pure nitrogen film was shown above in Fig. 4.

The reaction of dissociative recombination of $N_2^+$ with electron:

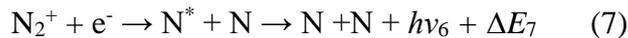
$$N_2^+ + e^- \rightarrow N^* + N \rightarrow N + N + h\nu_6 + \Delta E_7 \quad (7)$$

can be another source of excited atoms. For $N_2^+$ in the ground vibrational state, there are three dissociation limits of reaction (6): $N(^4S) + N(^2D)$, $N(^4S) + N(^2P)$, and $N(^2D) + N(^2D)$ [56]. Note, that reaction (7) yields "hot" N atoms with kinetic energy above 1 eV (1.06−3.45 eV) depending on the atomic state of products) [56] that is more than one order of magnitude larger than the binding energy $E_b$ per atom for solid Ne ($E_b$=26.5 meV [28]). Such energetic atoms produce defects in the immediate environment of the impurity which creates conditions for the formation of close pairs. Another recombination channel proceeds without dissociation and yields molecular emission:

$$N_2^+ + e^- \to N_2^* \to N_2 + h\nu_7 + \Delta E_8 \quad (8)$$

In this channel neutralization of molecular ions initially leads to population of high "Rydberg" states, which then quickly relax. This cascade eventually populates lower excited states such as $A^3\Sigma_u^+$ which decays with the emission of photons. Molecular bands of the Vegard-Kaplan transition are clearly seen in the TSL spectra of $N_2$-doped Ne matrix pre-irradiated with an electron beam (Fig. 6). A number of weak NIR bands in this figure (640, 706, 725 nm) stems from the transitions $3p_i \to 3s_k$ of Ne atoms which emerge via dissociative recombination of the neon self-trapped holes ($Ne_2^+$) with electrons:

$$Ne_2^+ + e^- \to Ne(3p) + Ne \to Ne + Ne + h\nu_8 + \Delta E_9 \quad (9)$$

Note, that in this relaxation channel, the formation of structural defects also occurs due to the repulsive interaction of the excited Ne atom with the environment, which leads to the formation of cavity with subsequent "plastic deformation". The appearance of additional structural defects that serve as traps for electrons contributes to the stabilization of positively charged species at low temperatures.

The most widely studied theoretically, are the $N_4^+$ cation and the neutral compound $N_4$ (Refs. [9-11] and references therein). The experimental studies of $N_4^+$ mentioned in [9–11] are also numerous. However, there are very few experimental reports on the observation of $N_4$. So, in [48, 49] the authors detected neutral $N_4$ starting from the known $N_4^+$ cation and using neutralization–reionization mass spectrometry (NRMS). They assigned their observations to the presence of a weakly bound open-chain $N_4$ complex with lifetime exceeding 1 μs. These observations were reproduced in the study [50] supported by high-level *ab initio* calculations. The authors suggested that the observed $N_4$ complex corresponds $^3N_4C_s$ open-chain structure which originates from a quartet $N_4^+$ excited state. Mass spectrometric evidence for the formation of $N_4$ as a result of the energy pooling reaction $N_2\,A^3\Sigma_u^+ + N_2\,A^3\Sigma_u^+$ was presented in [57]. There have also been several reports on the observation of a wide emission band at 360 nm in solid nitrogen [58] and nitrogen nanoclusters [59], which, based on a theoretical study [60] performed by the MCSCF method, was assigned to the $N_4$ ($D_{2h}$) structure of neutral $N_4$. According this study for an $N_4$ cluster of $D_{2h}$ symmetry the population of the $1B_{2u}$ state formed from the $A^3\Sigma_u^+$ and $B^3\Delta_u$ states of $N_2$ molecules may result in a radiative transition into the ground state $^1A_g$ at ≈ 3 eV. However, these theoretical calculations are inconsistent with the extended study of $N_4(D_{2h})$ excited states [61], which involved EOMCCSD calculations of vertical excitation energies and oscillator strengths for the lowest 20 singlet states of $N_4(D_{2h})$. By this study the excitation energy for the $1B_{2u}$ state is of 8.58 eV and the lowest excited state is $1B_{3u}$. The vertical absorption energy for the dipole-allowed transition from the ground state $A_g$ to the $1B_{3u}$ state is 1.58 eV and for the reverse radiative transition to the ground state – 1.55 eV respectively. The geometry the $1B_{3u}$ state is very similar to that of the ground state that results in large Frank-Condon factors for transitions between the two states. The effective barrier for dissociation from $N_4$ ($D_{2h}$) to $2N_2$ was estimated to be of 6.5 kcal mol$^{-1}$. The calculated emission energy of this transition turned out to be very close to the energy corresponding to the γ-line, but the experimental harmonic frequency is much lower than the lowest calculated one. This calls into question the interpretation of the γ-line as the radiative $1^1B_{3u} \to\ ^1A_g$ transition of the neutral compound $N_4$ of $D_{2h}$ configuration. Unfortunately, as far as we know, there are no data on the structure of $N_4^+$ in pure solid nitrogen. However, its structure in the Ne matrix was established [51, 52]. As it was mentioned above the structure of $N_4^+$ cation in this case corresponds to the ($D_{\infty h}$) open-chain configuration. In this regard, consider the data on open-chain configurations of $N_4$. As it was summarized in [11] high-level calculations found that the most stable $N_4$ isomer is an open chain. According to [62, 63] the $C_s$ structure of $N_4$ neutral appeared to be a minimum which lies 13.4 kcal mol$^{-1}$ below $N_4$ ($T_d$) [63]. A stable azidonitrene $C_s$ structure with a $^3A''$ ground state, and the singlet azidonitrene with a $C_s$ $^1A''$ state were reported in [64]. In both states, fragmentation giving $2N_2$ molecules needs to overcome a barrier height of about 55 kJ mol$^{-1}$. The authors of [50] considered the optimized ionic $N_4^+$ structures and the possible vertical neutralization products. The lowest-lying isomer with a relative G3 energy of 1403 kJ mol$^{-1}$ is the linear centrosymmetrical ($D_{\infty h}$) doublet ion with the $^2\Sigma_u^+$ ground state in accordance with previously obtained theoretical calculations. In this study authors also found an optimized doublet $N_4^+$ $C_s$ ion with a $^2A'$ state and the same G3 energy 1403 kJ mol$^{-1}$. Upon neutralization doublet ions can form singlet or triplet neutrals. The single-point energies which represent the final state in the vertical neutralization transitions have been obtained, i.e. in vertical neutralization the neutrals $N_4$ have been "frozen" in the geometry of the precursor ion. The respective G3 energies for the $^1N_4$ $D_{\infty h}$ and $^1N_4$ $C_s$ vertical neutrals were found to be 166 and 149 kJ mol$^{-1}$ above the dissociation products: $N_2$ ($X^1\Sigma_g^+$) + $N_2$ ($X^1\Sigma_g^+$), yielding the neutralization energies NEvs of 12.8 and 13.0 eV correspondingly. However, the authors were not able to locate the covalently bonded singlet neutral structure $^1N_4$ $C_s$ reported in [64]. Unfortunately, as far as we know, the energies of the excited states of neutrals of the $^1N_4$ $C_s$ and $^3N_4$ $C_s$ configurations have not been calculated. There are only data on the vibrational frequencies of different structures [50] (supporting information). The first conclusion that can be drawn from a comparison of the vibrational frequencies is that the registered structure is not a van der Waals complex (the measured frequency is 5 times higher than that characteristic of van der Waals structures). The experimental vibrational frequencies are close to the calculated ones [9, 50] for the open-chain structures $N_4$ $C_s$. However, because of incomplete data, the assignment of the γ-line and its satellites remains the subject of further experimental and theoretical studies.

**Summary**

The relaxation processes in solid nitrogen and in $N_2$-doped Ne matrix preirradiated with a low energy electron beam were studied using complex of emission spectroscopy methods: cathodoluminescence CL, developed by our group nonststionary luminescence NsL along with optical and current activation spectroscopy. Real-time correlated measurements of spectrally resolved thermally stimulated luminescence TSL and exoelectron emission were performed in the temperature range of the nitrogen α-phase existence and from 5 up to 16 K in the Ne matrix. Three emission bands were recorded in the NIR CL spectra of solid $N_2$: the so-called γ-line at 794 nm, and two satellites at 802, and 810 nm. The band at 810 nm was detected for the first time. It was found that these three bands which are characterized by similar behavior form molecular series with spacing between adjacent vibrational energy levels of the ground state of 125 cm$^{-1}$ and 123 cm$^{-1}$. The data obtained again raised the question of the identification of the γ-line, which in recent study [19] was attributed to the nitrogen anion N$^-$. Branched channels of electron-$N_n^+$ (n=1-4) interaction are discussed. Correlation of the γ-line with TSEE currents and recombination emission of O$^+$, $N_2^+$, and $N_4^+$ ions has been established, which points to its connection with the neutralization reaction. This conclusion was confirmed by NsL measurements. A new possible assignment of the γ-line as emission of tetranitrogen $N_4$ is discussed. Let us note that interest in research of polynitrogen compounds does not fade, which is evidenced by the appearance of new studies, in particular reviews [65-67].

**Acknowledgements**

The authors cordially thank Minh Nguyen, Tore Brinck, Alec Wodtke, Paul Mayer, Roman Boltnev and Vladimir Khmelenko for sharing relevant data and discussions.